\documentclass[runningheads]{llncs}
\usepackage[T1]{fontenc}
\usepackage{graphicx}

\usepackage{amsmath,amssymb,amsfonts}
\usepackage{algorithmic}
\usepackage{graphicx}
\usepackage{textcomp}
\usepackage[table]{xcolor}
\usepackage{xcolor}
\usepackage{booktabs} 
\usepackage{hyperref}
\usepackage{biblatex}
\usepackage{float}
\bibliography{Bib.bib}

\begin{document}
\title{Exploring Relationships Between Cryptocurrency News Outlets and Influencers' Twitter Activity and Market Prices}
\titlerunning{Cryptocurrency News and Influencers' Twitter Activity, and Market Prices}
%
\author{Meysam Alizadeh\inst{1} \and
Yasaman Asgari \inst{2}\ \and
Zeynab Samei \inst{3} \and   Sara Yari  \inst{1} \and \\ Shirin Dehghani  \inst{4} \and  Mael Kubli \inst{1} \and  Darya Zare \inst{5} \and Juan Diego Bermeo \inst{1} \and \\ Veronika Batzdorfer \inst{6} \and Fabrizio Gilardi \inst{1}} 
\authorrunning{M. Alizadeh et al.}

\institute{Digital Democracy Lab,
University of Zurich,
Zurich, Switzerland\\
\email{ \{syarim, alizadeh, kubli, jbermeo, gilardi\} @ipz.uzh.ch}
  \and
Digital Society Initiative
and Department of Mathematical Modeling and Machine learning,
University of Zurich,
Zurich, Switzerland
\email{yasaman.asgari@math.uzh.ch}\and
Department of Computer Science,
IPM,
Tehran, Iran
\email{zsamei@ipm.edu} \and Department of Computer Engineering,
Allameh Tabataba’i University,
Tehran, Iran
\email{sh.dehghani@atu.ac.ir} 
\and Department of Informatics,
University of Zurich,
Zurich, Switzerland
\email{daryazm@uzh.ch}
\and Institute for Sociology and Computational Social Science,
Karlsruhe Institute of Technology,
Karlsruhe, Germany
\email{vbatzdorfer@kit.de}}
\maketitle              
\begin{abstract}
Academics increasingly acknowledge the predictive power of social media for a wide variety of events and, more specifically, for financial markets. Anecdotal and empirical findings show that cryptocurrencies are among the financial assets that have been affected by news and influencers' activities on Twitter. However, the extent to which Twitter crypto influencer's posts about trading signals and their effect on market prices is mostly unexplored. In this paper, we use LLMs to uncover buy and not-buy signals from influencers and news outlets' Twitter posts and use a VAR analysis with Granger Causality tests and cross-correlation analysis to understand how these trading signals are temporally correlated with the top nine major cryptocurrencies' prices. Overall, the results show a mixed pattern across cryptocurrencies and temporal periods. However, we found that for the top three cryptocurrencies with the highest presence within news and influencer posts, their aggregated LLM-detected trading signal over the preceding 24 hours granger-causes fluctuations in their market prices, exhibiting a lag of at least 6 hours. In addition, the results reveal fundamental differences in how influencers and news outlets cover cryptocurrencies.
\keywords{Social Prediction\and NLP\and LLM \and Prompt Engineering \and
Time Series Analysis}
\end{abstract}

\section{Introduction}
Cryptocurrencies are a form of digital money that utilizes blockchain technology, a groundbreaking, decentralized, and encryption-based system that allows for the digital establishment of trust \cite{sockin2023decentralization}. Since the introduction of Bitcoin in 2009, the disruptive capabilities of cryptocurrencies have sparked significant growth and interest \cite{guidi2020blockchain}. This surge was mainly driven by media coverage highlighting the extraordinary returns on cryptocurrency investments, which led to a modern-day gold rush \cite{kraaijeveld2020predictive}. Moreover, the current international regulatory landscape for cryptocurrencies remains sparse, as they are mostly not yet recognized as a fully developed asset class \cite{sockin2023decentralization}. This lack of regulation, combined with their immense popularity and absence of institutional backing, contributes to the cryptocurrency market's extreme volatility and unpredictability, earning it the nickname “Wild West” \cite{kraaijeveld2020predictive}.

The instability of the cryptocurrency market is significantly driven by news and social media posts \cite{ante2023elon}. This dynamic is exacerbated by investors' challenges in verifying the accuracy of such information \cite{kraaijeveld2020predictive}. Given the relatively recent emergence of the cryptocurrency market, traditional media often lag in reporting developments, making social media a key information source for investors. Twitter (now re-branded as X), in particular, serves as a crucial platform for obtaining real-time updates and sentiment analysis on cryptocurrencies, allowing users to express their opinions and feelings \cite{kaminski2014nowcasting}. 

According to behavioral economics, emotions and sentiment can greatly influence individual actions and decision-making processes \cite{kahneman2003maps}. Previous research explored the effect of public Twitter sentiment on cryptocurrency prices \cite{kraaijeveld2020predictive}, Elon Musk's Twitter activity on the cryptocurrency market \cite{ante2023elon}, Twitter effect on stock market decisions during pandemics \cite{valle2022does}, and the predictive power of Twitter for Bitcoin price \cite{shen2019does} (see \cite{cano2023twitter} for a recent review). 

Given the abundance of Twitter data that captures the sentiments of cryptocurrency participants, the extent to which Twitter posts can help traders make informed decisions is mostly unexplored. Previous research on the association of Twitter posts with the cryptocurrency market suffers from at least two significant drawbacks. First, they mostly fail to account for fake accounts, such as bots and marketing campaigns. Indeed, previous research has shown that about 14\% of the active accounts in the cryptocurrency space are bots \cite{kraaijeveld2020predictive}. Second, the efficiency of the traditional sentiment analysis methods in capturing the true opinion of users is questionable, and more explicitly defined machine learning models that generate buy and sell signals \cite{kulakowski2023sentiment} may produce better results. 

In this paper, our goal is to determine whether LLM-detected trading signals from news outlets and crypto influencers' Twitter posts reflect cryptocurrency market dynamics. Our study utilizes text classification, network science methods, and times series analysis to analyze the retweet and co-mention networks. This structure consists of undirected links between cryptocurrencies that appear together in a tweet. In summary, we aim to answer two key research questions:
\begin{enumerate}
    \item How do the cryptocurrencies that are frequently co-mentioned together correlate with each other in their market prices? Do the links in the co-mention network reveal shared characteristics among cryptocurrencies, such as their technology or use cases?
    \item  Do the LLM-detected trading signals (buy or not-buy) of top cryptocurrencies obtained from news outlets and influencers' Twitter posts demonstrate a (lagged) correlation with their market prices?
\end{enumerate}


\section{Background}

\subsection{Effect of Sentiment on Financial Markets}

Kaplanski and Levy \cite{kaplanski2010sentiment} describe sentiment as any misconception that may lead to a deviation from the true fundamental value of an asset. Baker and Wurgler \cite{baker2007investor} note that the key factor distinguishing highly speculative assets is the difficulty and subjectivity involved in valuing them accurately. 
Fama et al. \cite{fama1969adjustment} argued that stock prices are unpredictable due to the erratic nature of news. According to Peterson \cite{peterson2016trading}, financial markets are considerably influenced by news, which in turn impacts sentiment. Traditionally, sentiment has been measured through investor surveys like those by the American Association of Individual Investors or Investor Intelligence. However, these methods are somewhat constrained by their dependence on achieving representative sample sizes. Recent advancements have seen a rise in the use of Natural Language Processing (NLP) to assess sentiment from news sources \cite{li2014news}. 

\subsection{Twitter Sentiment Analysis}

The effectiveness of using Twitter sentiment analysis to forecast financial markets is prominently illustrated in the study by Bollen et al. \cite{bollen2011twitter}, and further supported by Li et al. \cite{li2018more}. In their research, Li et al. \cite{li2018more} employ a Naive Bayes sentiment classifier and regression models to demonstrate that Tweets about stocks can predict daily stock returns. The findings also indicate that increases in Twitter activity follow higher market volatility from the previous day, positioning Twitter sentiment as both a predictor and a result of market movements. Furthermore, Mao et al. \cite{mao2011predicting} found that, unlike traditional investor sentiment, Twitter sentiment significantly predicts stock returns over the following 1–2 days. In general, the predictive power of Twitter sentiment for financial markets is generally observed to be the strongest between 1–4 days \cite{bollen2011twitter,mao2011predicting,zhang2011predicting}.

\subsection{Twitter and Cryptocurrency Market}

Social media has become the main source of information for cryptocurrencies. Several studies have analyzed user discussions on forums like Reddit and Bitcointalk.org (e.g. \cite{kim2017bitcoin,karalevicius2018using}). Many researchers recognize the short-term (1–7 days) and long-term (30–90 days) forecasting abilities of social media and news sentiment on Bitcoin’s price and trading volumes. The quantity of posts \cite{mai2015impacts} and Elon Musk's activity on Twitter \cite{ante2023elon} have been linked to the trading volume of Bitcoin \cite{mai2015impacts}. Similarly, Zou et al. \cite{zou2023prebit} proposed a multimodal model for Bitcoin extreme price movement prediction using Twitter. 

Twitter sentiment analysis has been used to predict changes in Bitcoin price. Georgoula et al. \cite{georgoula2015using} utilized a SVM to predict these fluctuations. 
Garcia and Schweitzer \cite{garcia2015social} applied a lexicon-based method along with a Vector Autoregressive (VAR) model and Granger-causality tests, discovering that rises in Twitter sentiment polarity often precede shifts in Bitcoin prices. Mai et al. \cite{mai2015impacts} performed intraday analysis and found Twitter posts effective in forecasting Bitcoin's hourly returns. Overall, the most effective number of lags is observed from 1–5 lags for interday analysis and 2–4 lags for intraday analysis \cite{mai2015impacts,karalevicius2018using}.

\subsection{Limitations of Current Literature}

The extant literature suffers from at least three major drawbacks: (1) much of the current literature is dedicated to Bitcoin, and other major coins have been neglected; (2) the vital role of the Twitter influencers has not been explored. Nearly all previous work has not distinguished between the original content produced by crypto experts and the content produced by lay users or even bots and marketing campaigns. 
(3) the meaning of ``Twitter sentiment" is rather vague as it may refer to emotion, stance, or buy or sell signals.


\section{Methods and Data}
\subsection{Twitter Data} \label{data:twitter}

We collect four different Twitter datasets. First, we used Twitter Academic API between 07/01/2022 and 14/01/2022 and between 01/03/2022 and 07/03/2022 and queried for `crypto' and a list of the top 200 cryptocurrencies to collect all tweets containing them (11.83M tweets posted by 2.1M unique users). Second, we manually searched for all authentic news outlets that have finance or cryptocurrency sections and have active Twitter accounts (74 accounts). Next, we collected all tweets posted by these 74 accounts between 01/01/2023 and 16/09/2023 (122,837 tweets). These two datasets serve for our influencer identification (see Section  \ref{inf_iden}) and model training purposes (see Section \ref{'twitter_pre} and \ref{signal_detection}). Our third and fourth datasets are created by continuously crawling Twitter data of the crypto-related influencers and news outlets between 10/06/2023 and 02/28/2024.

\subsection{Cryptocurrency Data}

We consider nine major cryptocurrencies according to their market capitalization. This includes Bitcoin (BTC),  Ethereum (ETH), Ripple (XRP), Solana (SOL), Dogecoin (DOGE), Binance coin (BNB), Ada (ADA), Polkadot (DOT), and Shiba Inu (SHIB). The financial information for these nine studied cryptocurrencies was obtained from CoinMarketCap (\href{https://coinmarketcap.com}{https://coinmarketcap.com}) from October 2023 to March 2024. 
CoinMarketCap is commonly used as a reference for cryptocurrency prices as it aggregates prices from numerous exchanges, providing a composite and more stable price estimate that avoids the biases of individual exchange prices. 

\subsection{Twitter Data Pre-processing}\label{'twitter_pre}

To prune out irrelevant tweets from our corpus, we trained and compared two classifiers and five LLMs on 500 annotated tweets. Tweets were classified as ``relevant'' if they were directly related to cryptocurrencies or related to economic or policy topics related to cryptocurrency market (e.g., US interest rate and cryptocurrency regulations). Out of the six classification models (\textit{CyptoBERT}, \textit{BERTweet}, \textit{GPT-3.5} trained with two distinct prompts, \textit{Random Forest with CryptoBERT Embedding}, and \textit{Random Forest and CryptoBERT-kk08 Embedding}), we achieved the maximum accuracy of 0.94 on a 20\% unseen out-of-sample test data for GPT-3.5 with optimized prompt. Therefore, for all four datasets we used GPT-3.5 with an optimized prompt to exclude irrelevant tweets.

\begin{table*}[h!]
     \centering
     \tiny
     \resizebox{\textwidth}{!}{
         \begin{tabular}{llll}
             \specialrule{\heavyrulewidth}{\abovetopsep}{\belowbottomsep}
             model & balancing method & accuracy & f1-score \\
             \hline
             CoinBERT & unbalance & 92\% & 91\% \\
             BERTweet & gpt3.5 augmentation & 92\% & 91\%\\
             \textbf{CoinGPT} & - & \textbf{96}\% & \textbf{94}\%\\
             GPT-4 & - & 93\% & 91\%\\
             Random forest+cryptobert embedding & smote & 88\% & 82\%\\
             Random forest+cryptobert-kk08 embedding & smote & 86\% & 81\%\\
             \hline
         \end{tabular}
     }
     \caption{\label{relevant_tweets}Comparing the accuracy of various classifiers for pruning irrelevant tweets.}
\end{table*}
\subsection{Influencer Identification}\label{inf_iden}

We use four different methodologies to identify crypto-related influencers on Twitter. First, we used our nearly 12 million tweets dataset to construct a retweet network. Following Kwak et al. \cite{kwak2010twitter}, we calculate three centrality metrics, including \textit{PageRank}, \textit{Betweenness}, and \textit{Closeness} scores for all users within the retweet network. From each centrality measure, we select the top 1000 accounts. After combining these lists, we obtained 1,283 users. 


Second, we examined the marketing campaigns of 10 coins that were launched in 2023. 
Through the retweet network analysis, we identified the influential accounts. 
This led to identifying 87 new influencers based on their \textit{PageRank} scores. Third, we crafted a list of 50 non-major cryptocurrency coins and their related keywords and collected all tweets posted between 01/01/2024 to 01/02/2024. 
This process yielded 11,408 users who were potential influencers. Finally, we used LunarCrush (\href{https://lunarcrush.com/}{https://lunarcrush.com/}) website and scraped the list of top 1000 influencers for each of the leading 100 coins. After removing mutual influencers, we obtained 36,311 unique users from LunarCrush. 

        


After combining these four lists and excluding duplicates, we collected user profile information (bio descriptions and follower counts) and the 10 most recent tweets for each account. Users with irrelevant descriptions, fewer than 5000 followers, no tweets in the last 3 months, and users with an average engagement of fewer than 200 in their last 10 tweets were excluded. 
Overall, we obtained 2,687 cryptocurrency influencers on Twitter.





        
    

\subsection{Trading Signal Detection}\label{signal_detection}

We use CryptoBERT\cite{kulakowski2023sentiment} to detect buy, sell, and neutral signals from the text of the influencers' tweets. CryptoBERT is trained on a corpus of over 3.2 million crypto-related social media posts. 
We compare its performance with \textit{Random Forests}, \textit{Logistic Regression}, \textit{SVM}, and \textit{XGBoost}. Our feature engineering includes content features such as top unigrams, top bigrams, top hashtags, top mentions, number of mentions, number of emojis, and number of hashtags. Considering the previous findings on the ability of ChatGPT \cite{gilardi2023chatgpt} and open-source LLMs \cite{alizadeh2023open} in text annotation tasks, we included GPT-4, GPT-3.5, LlaMa-2 (7b), LlaMa-2 (70b), FLAN-T5 (L), and FLAN-T5 (XL) in our comparison. We randomly selected 600 tweets from influencers and 200 tweets from news outlets' datasets, and labeled them for buy, sell, and neutral signals (Table \ref{labled_tweets}). As can be seen in Table \ref{labled_tweets}, Since the frequency of `sell (bearish)' signal in the influencer's data is low (5.5\% of total labeled data), we combined the `sell' and `neutral' classes into one class and called it the `not-buy' class. 

\begin{table}[h!]
    \centering
    \footnotesize
        \caption{\label{labled_tweets}The distribution of signals across annotated datasets.}
    \begin{tabular}{llc}
        \specialrule{\heavyrulewidth}{\abovetopsep}{\belowbottomsep}
        Dataset &\hspace{1cm}Class names &\hspace{1cm}$\%$ samples \\
        \hline
        Twitter Influencers &\hspace{1cm} Buy (Bullish) & \hspace{1cm}56.8 \\
        &\hspace{1cm} Sell (Bearish) &\hspace{1cm}5.50 \\
        &\hspace{1cm} Neutral &\hspace{1cm}37.7 \\
        \hline
        News Outlets &\hspace{1cm} Buy (Bullish) &\hspace{1cm}30.5 \\
        &\hspace{1cm} Sell (Bearish) &\hspace{1cm}25.5 \\
        &\hspace{1cm} Neutral &\hspace{1cm}44.0 \\
        \specialrule{\heavyrulewidth}{\abovetopsep}{\belowbottomsep}
    \end{tabular}

\end{table}

\begin{figure}[h!]
   \centering
   \footnotesize
   \includegraphics[width=0.8\textwidth ]{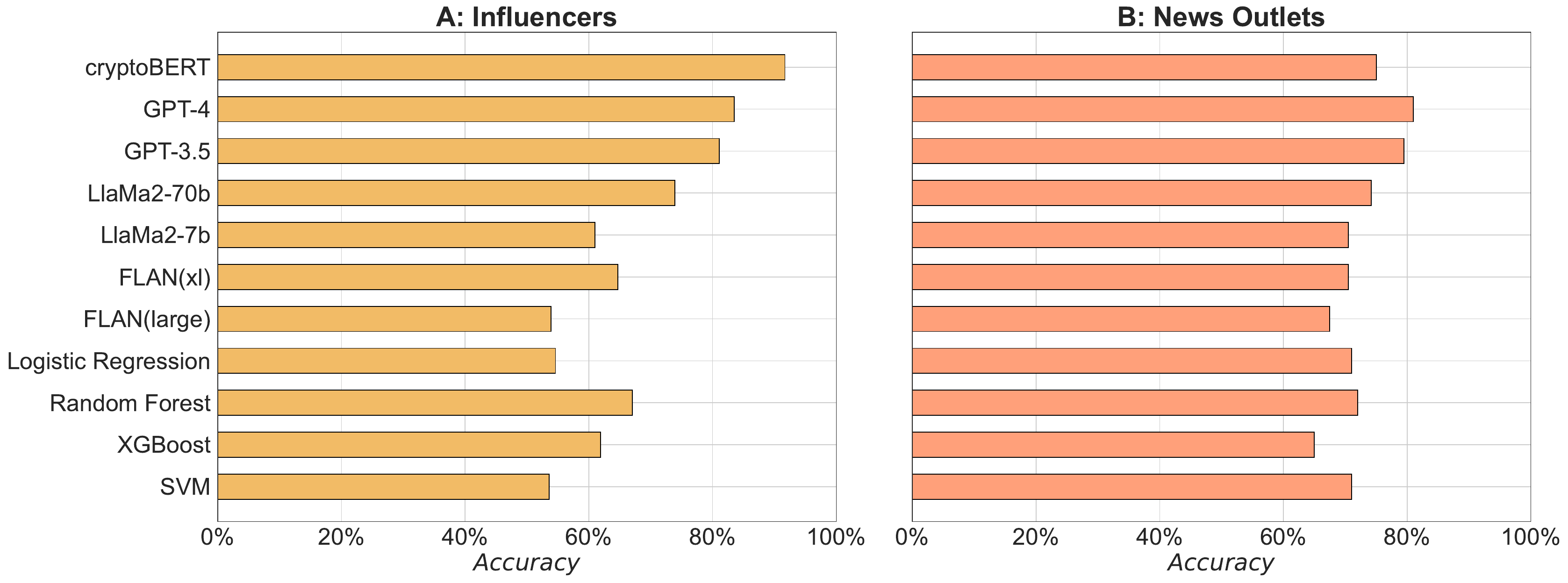}
  \caption{\footnotesize Choosing the best AI model for the trading signal detection task.}
    \label{2class_signal}
\end{figure}

In Figure \ref{2class_signal}, we see the CryptoBert model outperforms all other models for the influencers dataset. 
However, for the news outlets' data, GPT-4 and GPT-3.5 are the two best-performing models and slightly outperform CryptoBERT. Since CryptoBERT's performance is strong across both datasets, and it is open-source (and thus more ethical for academic purposes) and free to use, we selected it as our final choice.

\subsection{Constructing Co-mention Network}

In a co-mention network, nodes represent individual cryptocurrency coins, and edges symbolize the two coins that were mentioned together in a tweet. 
The network is undirected and weighted, characterized by symmetric adjacency matrix $A$ with entries $w_{ij} \in \mathbb{N} $ corresponding to the frequency of co-mentions between coins \(i\) and \(j\). 



\subsection{Time series processing}\label{timeseries_label}
Time series data consist of a sequence of values or observations recorded over time. Mathematically, a time series can be defined as a set of ordered observations \( X = \{x_1, x_2, \ldots, x_n\} \), where each \( x_i \) corresponds to the value of the series at the \( i^{th} \) time point.

The Granger causality test assesses whether one time series can predict another, a method commonly used to investigate economic time series interactions. The null hypothesis states that \( X \) does not cause \( Y \) in the Granger sense. The procedure involves formulating a vector autoregression (VAR) model for the time series under study. For two variables \(X\) and \(Y\), the model can be specified as:
\begin{align}
    Y_t &= \alpha + \sum_{i=1}^k \beta_i Y_{t-i} + \sum_{i=1}^k \gamma_i X_{t-i} + \epsilon_t\\
    X_t &= \delta + \sum_{i=1}^k \phi_i X_{t-i} + \sum_{i=1}^k \psi_i Y_{t-i} + \eta_t
\end{align}

Here, \(\alpha\) and \(\delta\) represent constant terms, \(\beta_i, \gamma_i, \phi_i,\) and \(\psi_i\) are the coefficients, and \(\epsilon_t\) and \(\eta_t\) denote the error terms of the model. The parameter \(k\) signifies the number of lagged observations in the model. An F-statistic is calculated to evaluate the null hypothesis. 
Granger causality testing identifies patterns of lagged correlation but does not confirm causality. This concept is akin to how observing cloud cover, which often precedes rain, can help predict rain but does not cause it.

Cross-correlation evaluates the degree to which two series are correlated at different time lags \( k \). If we denote another time series as \( Y = \{y_1, y_2, \ldots, y_n\} \), the cross-correlation \( \gamma \) at lag \( k \) can be computed using \eqref{time_series}.
The numerator of the formula represents the covariance between the \( x \) series shifted by \( k \) units in time and the \( y \) series. The denominator normalizes this value, ensuring that the cross-correlation coefficient \( \gamma \) lies between -1 and 1, inclusive. The sign and magnitude of \( \gamma \) at different values of \( k \) can reveal the leading and lagging relationships between the two-time series. 
\begin{equation}
   \gamma = \frac{\sum_{i} (x_{i+k} - \bar{X})(y_i - \bar{Y})}{\sqrt{\sum_{i}(x_{i+k} - \bar{X})^2} \sqrt{\sum_{i}(y_i - \bar{Y})^2}} 
   \label{time_series}
\end{equation}



\section{Results}
\subsection{Description of Influencers and News Outlets Twitter Data}

\begin{figure*}
   \centering
   \footnotesize
   \includegraphics[ width=0.8\textwidth , keepaspectratio]{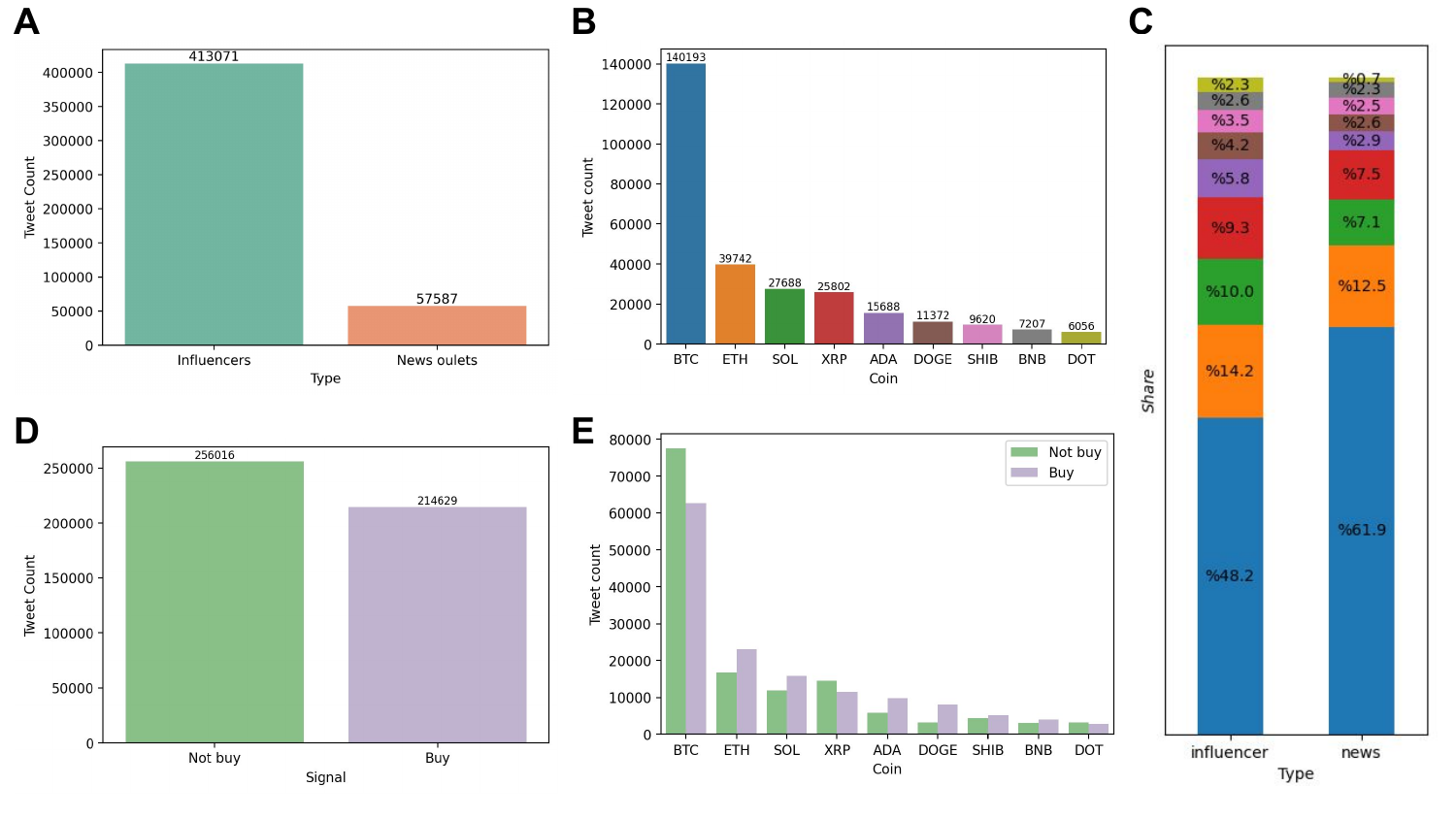}
    \caption{\footnotesize \textbf{Exploratory Analysis}: (A) User Type Distribution (B) Frequency of Tweets by Major Coin, with Bitcoin as the Most Mentioned (C) Comparison of 9-major coins' mentions in News Outlets versus Influencer Tweets (D) Signal Distribution (E) 'Not Buy' signals predominant for BTC, XRP, and DOT, while 'Buy' Signals are more common for ETH, SOL, ADA, DOGE, SHIB, and BNB.}
    \label{figure2}
\end{figure*}

The final Twitter dataset used in this experiment contains 470,658 English tweets from influencers and news outlet's Twitter accounts, between 10/06/2023 and 02/28/2024.
Tweets written by influencers dominate the dataset with 413,071 tweets (87.8\%), while news outlets account for 57,587 tweets (12.2\%). This underscores the pivotal role of influencers in shaping discussions and influencing trading decisions within the cryptocurrency market (Figure \ref{figure2}-A). Moreover, Bitcoin appears in 340,193 tweets (34.8\%), followed by Ethereum and Solana with 39,742 (9.9\%) and 27,688 (6.9\%) tweets, respectively (Figure \ref{figure2}-B). 
Figure \ref{figure2}-C illustrates the coin mention share distribution, highlighting structural variations between influencers and news outlets. An interesting observation here is that the share of Bitcoin mentions in news outlets' tweets (61.9\%) is higher than influencers' tweets (48.2\%). 

Figure \ref{figure2}-D shows the dataset predominantly contains `not buy' signal, representing 54.39\% (256,016 tweets), while `buy' signals account for 45.60\% (214,629 tweets). Figure \ref{figure2}-E details the signal distribution across nine major coins. BTC, XRP, and DOT predominantly feature `not buy' signals, whereas ETH, SOL, ADA, DOGE, SHIB, and BNB mainly show `buy' signals. Notably, DOGE has 2.53 times more `buy' than `not buy' signals, the highest ratio amongst all examined coins.

\subsection{Analyzing the cryptocurrencies co-mention network}

\subsubsection{Exploratory Analysis of the Co-Mention Network}
\begin{figure}[!ht]
   \centering
   \footnotesize   \includegraphics[width=0.8\textwidth,keepaspectratio]{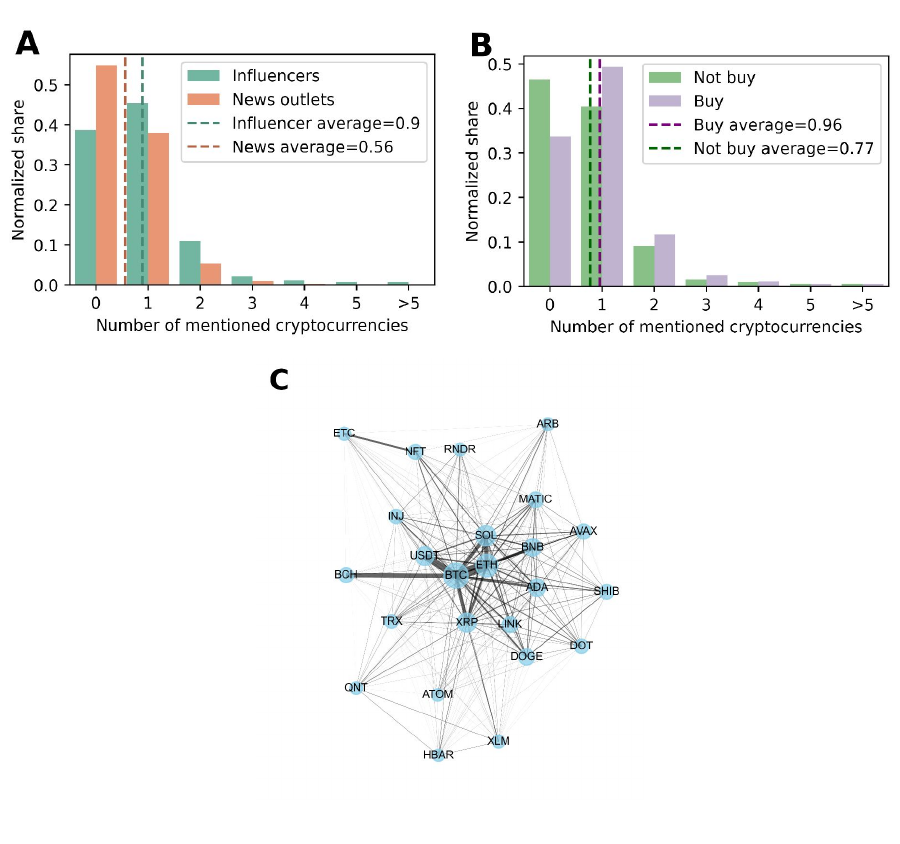}
  \caption{\footnotesize Structure of the cryptocurrencies co-mention network: (A) Influencers mention more coins in their tweets than news outlets do. (B) Tweets signaling a buy mention more coins than tweets not signaling a buy. (C) The co-mention network of top 24 coins. There exists a notable connection between NFT and ETC.}
    \label{figure1}
\end{figure}

This study focuses on the co-mention network of the top 100 coins according to their market capitalization. The dataset includes tweets mentioning between zero (generally discussing the cryptocurrency financial market) to over 40 coins, with the most common scenario involving single-coin mentions (44.4\% of tweets). We also examine the differences in how influencers and news outlets reference cryptocurrencies. According to Figure \ref{figure1}-A, influencers tend to mention more coins per tweet (0.9) compared to news outlets (0.56). This suggests that influencers may discuss a broader range of coins, likely to engage a larger audience or to demonstrate their knowledge across various market segments.

Furthermore, Figure \ref{figure1}-B reveals distinct characteristics in tweets signaling `buy' versus `not buy' intentions. Tweets encouraging buying typically mention at least one specific coin, suggesting a direct investment recommendation. In contrast, `no buy' tweets often do not specify any coins and focus more on general market conditions or cautions. The average number of coins mentioned in `buy' tweets is higher (0.96) compared to `not buy' tweets (0.77), showing a more targeted approach to investment suggestions.

We reveal key patterns in the co-mention network using the k-core method to filter the network, where nodes must have connections involving at least 1\% of the total degree sum. Figure \ref{figure1}-C shows this filtered k-core network of 24 nodes. The edge thickness represents the edge weight, while the node size indicates the node's degree of centrality. 
The data highlights several key connections: BTC to ETH with 14,062 connections (5.16 \%), BTC to USDT (3.54 \%), BTC to BCH (2.37 \%), ETH to SOL (2.14 \%), BTC to XRP (1.81 \%), BTC to SOL (1.79 \%), BTC to ADA (1.25 \%), and XRP to ETH (1.18 \%). Bitcoin (BTC) and Ethereum (ETH) are frequently featured, with a notable connection between Ethereum Classic (ETC) and Non-Fungible Tokens (NFT).

\subsubsection{Interplay between the co-mention network and cryptocurrency attributes}
Next, We investigate whether the connections in the co-mention network indicate common features among cryptocurrencies, such as their technology or applications. Following Mungo et al \cite{mungo2024cryptocurrency}, we utilize tags from Coinmarketcap that describe various cryptocurrencies' primary attributes. These tags detail each cryptocurrency's blockchain technology and information about its ecosystem, such as whether the cryptocurrency is built on an independent or existing blockchain and is involved in decentralized finance (DeFi) projects. Additionally, the tags describe the specific functions and utilities of the cryptocurrencies, for instance, their use in distributed storage, as fan tokens, or as digital stores of value like digital gold.

Following the methodology suggested by Mungo et al. (\cite{mungo2024cryptocurrency}), we assign a vector \(x_i\) to each cryptocurrency, where each tag \(j\) is represented as \(x_{i,j} = 1\) if it applies to the cryptocurrency and \(x_{i,j} = 0\) otherwise. We then calculate the cosine similarity in the characteristics space between each pair of cryptocurrencies to create a similarity matrix. Next, we compute a Pearson correlation between the weighted adjacency and similarity matrices. The correlation coefficient of \(0.1735\) with a p-value of \(<0.001\) suggests that the patterns observed in the co-mention network are only slightly correlated with the coin characteristics, and there are other mechanisms run by the influencers and news outlets on the mentioning patterns within the social media. 

\subsubsection{Interplay between the co-mention network structure and cryptocurrency price returns}
We analyze pairs of coins in the co-mention network that constitute at least 1\% of the total connections. From the initial set of 24 cryptocurrencies considered, we exclude two: NFT and USDT. NFTs, or non-fungible tokens, differ from cryptocurrencies as they do not have standard prices listed on platforms like CoinMarketCap due to the unique nature of each NFT, which contrasts with fungible assets like Bitcoin or Ethereum where each unit is interchangeable. Additionally, USDT is excluded because it is a stablecoin. 

\begin{figure}
   \centering
   \includegraphics[width=0.55\textwidth]{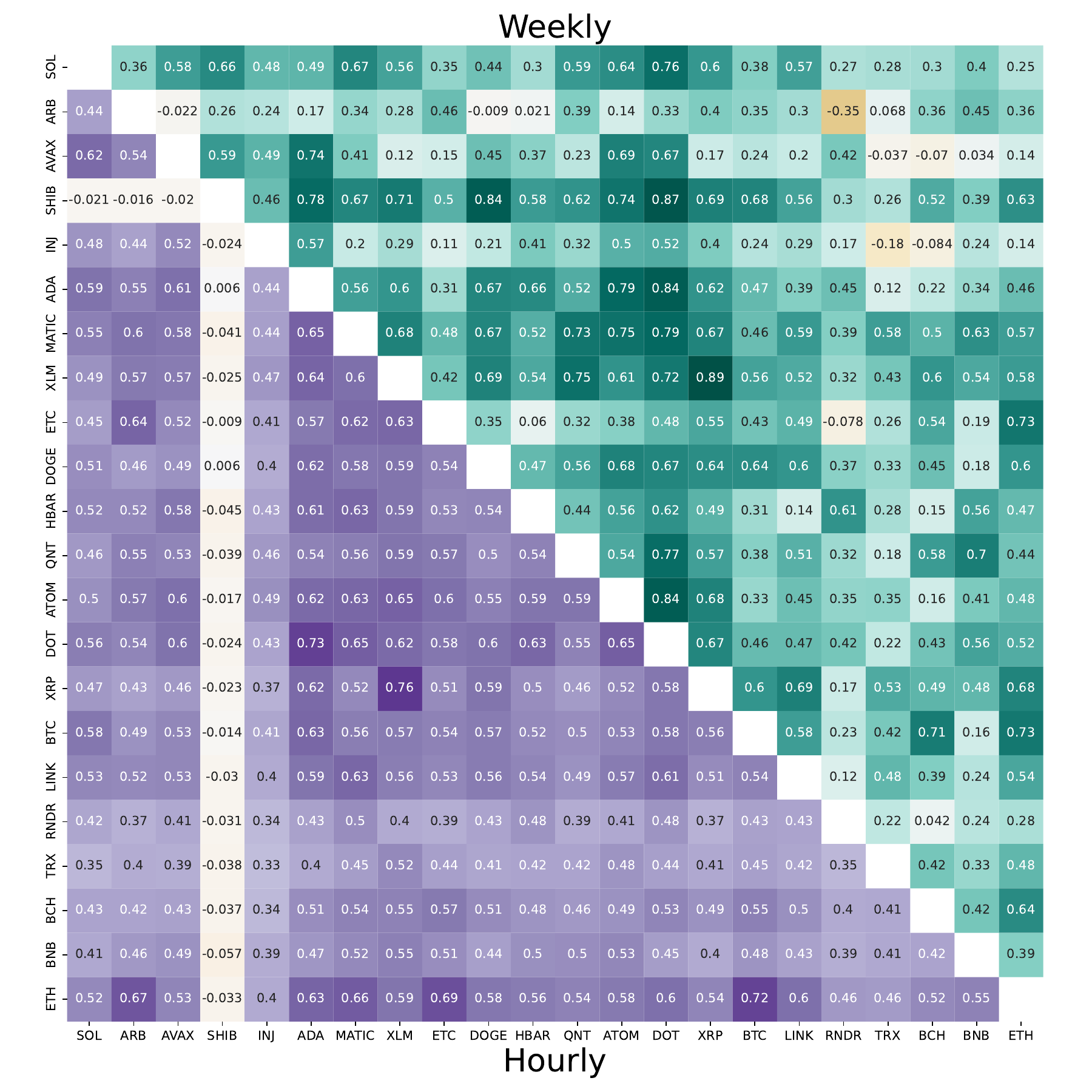}
 \caption{Correlation matrix for log returns of prices for pairs of coins in the network holding at least \(1\%\) degree share. Upper triangle: Long-term weekly cross-correlation analysis (\(k=0\)). Lower triangle: Short-term hourly cross-correlation analysis (\(k=0\)).}
    \label{figure3}
\end{figure}

We construct a \(22\times22\) correlation matrix, denoted as \(\mathbf{C}\), in which each cell \(\mathbf{C}_{ij}\), represents the cross-correlation at zero lag between coins \(i\) and \(j\). This correlation is calculated differently depending on the time scale: for pairs where \(i>j\), we use hourly log returns, while for \(i<j\), we use log returns for mean weekly prices. This dual-time approach allows us to capture both short-term and long-term interdependencies between cryptocurrencies. In Figure \ref{figure3}, the lower triangle displays the correlations computed from hourly data, and the upper triangle shows the correlations from weekly data. Prior to conducting the correlation analysis, an Augmented Dickey-Fuller test confirmed that all price log return series are stationary, each with a p-value of \(<0.001\). All the correlation values are statistically significant with p-value\(<0.001\).

The correlation values for short-term dynamics consistently exceed 0.3 across all pairs except for SHIB. However, for the long-term test, several pairs show zero or negative correlations. The highest correlation is 0.89 for the long term between XLM and XRP and 0.78 for the short term. Following closely behind for short-term dynamics are DOT-ADA (0.73) and BTC-ETH (0.72). 
Conversely, the strongest correlations for long-term dynamics are noted for SHIB-DOT (0.87) and DOT-ATOM (0.84). The lowest short-term correlation values are predominantly between SHIB and all other coins, with the lowest long-term correlation being between ARB and RNDR (-0.35).

We hypothesize that a higher weight between pairs in the co-mention network leads to higher similarity in the logarithmic returns of their hourly and mean weekly prices. To test this hypothesis, we perform a Pearson correlation analysis between the adjacency matrix of the co-mention network and the correlation matrix \(\mathbf{C}\) for the log returns of hourly and weekly prices. The resulting correlation coefficient for hourly prices is 0.154, with a p-value of 0.019, and for weekly prices is 0.16, with a p-value of 0.015. This indicates that while there is a statistically significant relationship, the effect size is relatively small. This suggests that factors other than social media co-mentions likely play a more significant role in influencing financial market movements. Examples of these influential factors can be shared with Investors because investors often move in herds; thus, if a significant investor adjusts their portfolio across several assets, it could lead to simultaneous price movements in those assets. Similar Technological Foundations or common applications could also be another factor. 

\subsection{Correlation of LLM-detected signals and cryptocurrency prices}
In this section, we examine the existence of Granger causality and lagged cross-correlation between the market prices of nine major cryptocurrencies and their associated social signals. Our goal is to identify the presence of these relationships and the specific lags where they become significant. This analysis also investigates if the optimal lag differs among cryptocurrencies, indicating that the interplay between market dynamics and social signals may be coin-specific.

The 24-hour Twitter social signal for each cryptocurrency \(i\) at time \(t\) is quantified as \eqref{social-signals}. An increase in \(\text{SS}^i_t\) is hypothesized to correlate with a rise in the cryptocurrency's price.
\begin{equation}
    \text{SS}^i(t )=\left(\frac{1+N^i_{\text{Buy}}(t)}{1+N^i_{\text{Not buy}}(t)}\right)
    \label{social-signals}
\end{equation}
We can also include information about the cryptocurrency market (Tweets related to crypto but without mentioning any coin) in our analysis because the market status influences prices simultaneously.
\begin{equation}
    \text{SS}^i_{\text{Crypto}}(t)=\left(\frac{1+N^i_{Buy}(t)+N_{\text{Buy}}^{\text{ Crypto}}(t)}{1+N^i_{\text{Not buy}}(t)+N_{\text{Not buy}}^{\text{ Crypto}}(t)}\right)
    \label{social-signals-crypto}
\end{equation}
The log return of the social signal for cryptocurrency \(i\) from time \(t-1\) to \(t\) is denoted by \(r^i_{\text{SS}}(t)\). Similarly, cryptocurrency prices exhibit non-stationary behavior, and the log-returns of the prices are denoted by \(r^i_{\text{CP}}(t)\).

Table \ref{granger} illustrates a Granger causality analysis between Twitter social signals, $r^i_{\text{SS}_{\text{crypto}}}(t)$, and price changes of nine major cryptocurrencies,$r^i_{\text{CP}}(t)$, highlighting how these signals may predict price movements. 
BTC and ETH demonstrate significant predictive correlations at multiple hourly lags of less than 10 hours, whereas DOGE exhibits signs of correlation after 10 hours, and SOL for the entire 24 hours, suggesting a robust effect of influencers and news outlets on coins prices. In contrast, ADA, XRP, and DOT show significant Granger causal links predominantly within the first three hours, indicating a rapid market response to new social data. Conversely, BNB exhibits week predictive lags, implying that Twitter data does not hugely influence price changes. SHIB reveals predictive power specifically at 21-22 hour lags, showcasing delayed market reactions to social signals. DOGE and SOL also present varied predictive lags, reflecting their community-driven market dynamics. This analysis underscores the diverse ways different cryptocurrencies respond to social media, driven by unique investor behaviors and market conditions.

\begin{table}[!h]
\tiny
    \centering
\begin{tabular}{lccccccccc}\hline
 & ADA & BNB & BTC & DOGE & DOT & ETH & SHIB & SOL & XRP \\
\hline
1H & \cellcolor{orange!25}0.040 & 0.111 & \cellcolor{cyan!25}0.001 & \cellcolor{orange!25}0.017 & \cellcolor{orange!25}0.043 & \cellcolor{cyan!25}0.002 & 0.812 & \cellcolor{cyan!25}0.001 & \cellcolor{orange!25}0.021 \\
2H & \cellcolor{lime!25}0.077 & \cellcolor{lime!25}0.080 & \cellcolor{cyan!25}0.003 & \cellcolor{orange!25}0.025 & \cellcolor{lime!25}0.085 & \cellcolor{cyan!25}0.003 & 0.898 & \cellcolor{cyan!25}0.001 & \cellcolor{lime!25}0.053 \\
3H & \cellcolor{orange!25}0.047 & \cellcolor{lime!25}0.090 & \cellcolor{cyan!25}0.001 & \cellcolor{cyan!25}0.006 & \cellcolor{lime!25}0.063 & \cellcolor{cyan!25}0.003 & 0.978 & \cellcolor{cyan!25}0.001 & \cellcolor{orange!25}0.047 \\
4H & \cellcolor{lime!25}0.063 & \cellcolor{lime!25}0.083 & \cellcolor{cyan!25}0.002 & \cellcolor{orange!25}0.013 & 0.103 & \cellcolor{cyan!25}0.008 & 0.961 & \cellcolor{cyan!25}0.003 & \cellcolor{lime!25}0.070 \\
5H & 0.100 & 0.118 & \cellcolor{cyan!25}0.001 & \cellcolor{orange!25}0.013 & 0.185 & \cellcolor{orange!25}0.010 & 0.972 & \cellcolor{cyan!25}0.004 & 0.101 \\
6H & \cellcolor{lime!25}0.091 & 0.180 & \cellcolor{cyan!25}0.001 & \cellcolor{orange!25}0.020 & 0.215 & \cellcolor{cyan!25}0.006 & 0.887 & \cellcolor{cyan!25}0.003 & \cellcolor{lime!25}0.058 \\
7H & \cellcolor{lime!25}0.094 & 0.230 & \cellcolor{cyan!25}0.002 & \cellcolor{orange!25}0.031 & 0.297 & \cellcolor{orange!25}0.012 & 0.236 & \cellcolor{cyan!25}0.006 & \cellcolor{lime!25}0.072 \\
8H & 0.104 & 0.300 & \cellcolor{cyan!25}0.004 & \cellcolor{orange!25}0.015 & 0.353 & \cellcolor{orange!25}0.015 & 0.324 & \cellcolor{cyan!25}0.005 & \cellcolor{lime!25}0.065 \\
9H & 0.144 & 0.383 & \cellcolor{cyan!25}0.006 & \cellcolor{orange!25}0.021 & 0.435 & \cellcolor{orange!25}0.024 & 0.320 & \cellcolor{cyan!25}0.008 & \cellcolor{lime!25}0.091 \\
10H & 0.187 & 0.466 & \cellcolor{cyan!25}0.008 & \cellcolor{cyan!25}0.001 & 0.266 & \cellcolor{orange!25}0.023 & 0.412 & \cellcolor{orange!25}0.011 & 0.141 \\
11H & 0.209 & 0.560 & \cellcolor{orange!25}0.012 & \cellcolor{cyan!25}0.003 & 0.334 & \cellcolor{orange!25}0.025 & 0.490 & \cellcolor{orange!25}0.014 & 0.105 \\
12H & 0.230 & 0.640 & \cellcolor{orange!25}0.014 & \cellcolor{cyan!25}0.004 & 0.414 & \cellcolor{orange!25}0.019 & 0.523 & \cellcolor{cyan!25}0.009 & 0.127 \\
13H & 0.282 & 0.647 & \cellcolor{orange!25}0.023 & \cellcolor{cyan!25}0.006 & 0.429 & \cellcolor{orange!25}0.026 & 0.438 & \cellcolor{orange!25}0.011 & 0.158 \\
14H & 0.339 & 0.655 & \cellcolor{orange!25}0.034 & \cellcolor{cyan!25}0.009 & 0.509 & \cellcolor{orange!25}0.041 & 0.275 & \cellcolor{cyan!25}0.005 & 0.193 \\
15H & 0.296 & 0.687 & \cellcolor{orange!25}0.046 & \cellcolor{cyan!25}0.008 & 0.545 & \cellcolor{lime!25}0.051 & 0.330 & \cellcolor{cyan!25}0.007 & 0.214 \\
16H & 0.318 & 0.735 & \cellcolor{lime!25}0.057 & \cellcolor{cyan!25}0.009 & 0.469 & \cellcolor{orange!25}0.042 & 0.330 & \cellcolor{cyan!25}0.003 & 0.253 \\
17H & 0.375 & 0.793 & \cellcolor{lime!25}0.078 & \cellcolor{cyan!25}0.008 & 0.493 & \cellcolor{lime!25}0.062 & 0.241 & \cellcolor{cyan!25}0.004 & 0.292 \\
18H & 0.435 & 0.843 & 0.111 & \cellcolor{orange!25}0.013 & 0.521 & \cellcolor{lime!25}0.065 & 0.215 & \cellcolor{cyan!25}0.002 & 0.263 \\
19H & 0.366 & 0.794 & \cellcolor{lime!25}0.099 & \cellcolor{cyan!25}0.009 & 0.580 & \cellcolor{lime!25}0.078 & 0.277 & \cellcolor{cyan!25}0.002 & 0.261 \\
20H & 0.192 & 0.795 & 0.116 & \cellcolor{cyan!25}0.006 & 0.502 & \cellcolor{lime!25}0.059 & 0.272 & \cellcolor{cyan!25}0.001 & 0.105 \\
21H & 0.217 & 0.808 & 0.137 & \cellcolor{cyan!25}0.006 & 0.543 & \cellcolor{lime!25}0.079 & \cellcolor{orange!25}0.046 & \cellcolor{cyan!25}0.001 & 0.126 \\
22H & 0.263 & 0.823 & 0.158 & \cellcolor{cyan!25}0.009 & 0.591 & \cellcolor{lime!25}0.090 & \cellcolor{orange!25}0.042 & \cellcolor{cyan!25}0.001 & 0.161 \\
23H & 0.291 & 0.861 & 0.166 & \cellcolor{orange!25}0.013 & 0.633 & 0.110 & \cellcolor{lime!25}0.051 & \cellcolor{cyan!25}0.002 & 0.150 \\
24H & 0.398 & 0.859 & 0.355 & \cellcolor{cyan!25}0.007 & 0.386 & \cellcolor{lime!25}0.060 & 0.169 & \cellcolor{cyan!25}0.001 & 0.112 \\
\hline\\
\end{tabular}
    \caption{Granger causality analysis between Twitter social signal  $r^i_{\text{SS}_{\text{crypto}}}(t)$ and prices changes of 9 major cryptocurrencies $r^i_{\text{CP}}(t)$. Blue color shows \(p-\)value \(<0.01\), orange shows \(p-\)value \(<0.05\) and green, \(p-\)value \(<0.1\). }
    \label{granger}
\end{table}

\begin{table*}[h]
\tiny
\centering
\caption{Highest correlation coefficient with associated lags for 9 major coins with different formulas}

\begin{tabular}{c*{10}{|@{\hspace{1pt}}c@{\hspace{1pt}}}}

\hline \rule{0pt}{9pt}
Price& Social signal&  ADA & BNB &  BTC & DOGE & DOT & ETH & SHIB & SOL& XRP \\[3pt] \hline
\rule{0pt}{12pt}
\(\text{CP}^i(t)\) & \(\text{SS}^i(t)\) & \(0.09^{(7D)}\)& \(-0.15^{(5D)}\) & \(-0.15^{(6D)}\) & \(-0.25^{(2H)}\) & \(-0.35^{(3D)}\) &\( 0.23^{(1H)}\) & \(-0.05^{(6D)}\) & \(0.31^{(6D)}\) & \(0.06^{(2H)}\) \\[10pt] \hline   \rule{0pt}{12pt}
\(\text{CP}^i(t)\) &  \(\text{SS}_{\text{crypto}}^i(t)\)  & \(0.25{(0H)}\)& \(0.06^{(0H)}\) & \(-0.03^{(6D)}\) & \(0.22^{(1H)}\) & \(0.19^{(0H)}\) &\( 0.23^{(0H)}\) & \(0.25^{(0H)}\) & \(0.33^{(10H)}\) & \(0.44^{(2H)}\) \\[10pt] \hline   \rule{0pt}{12pt}
\(r^i_{\text{CP}}(t)\) & \(r^i_{\text{SS}}(t)\)  & \(0.03^{(6D)}\)& \(0.03^{(18H)}\) & \(0.08^{(0H)}\) & \(0.04^{(0H)}\) & \(0.05^{(0H)}\) &\( 0.07^{(1H)}\) & \(0.03^{(20H)}\) & \(0.04^{(1H)}\) & \(0.04^{(8H)}\) \\[10pt] \hline \rule{0pt}{12pt}
\(r^i_{\text{CP}}(t)\) & \(r^i_{\text{SS, crypto}}(t)\) & \(0.04^{(0H)}\)& \(0.05^{(0H)}\) & \(0.08^{(0H)}\) & \(0.05^{(0H)}\) & \(0.04^{(0H)}\) &\( 0.06^{(1H)}\) & \(0.04^{(7H)}\) & \(0.06^{(1H)}\) & \(0.04^{(20H)}\) \\[10pt] \hline 

\end{tabular}
\label{table}
\end{table*}

Table \ref{table} illustrates the highest cross-correlation values between each cryptocurrency's price and its LLM-detected trading signal aggregated on influencers and news outlets using hourly (24 hours) and daily lags (up to 7 days) to capture short-term and long-term market behaviors. Each row represents a different computation formula for the price and social trading signals, and each cell below the cryptocurrency names columns shows the highest correlation value for a cryptocurrency with the associated lag.  In general, we can see that the coefficients for price are much larger than the coefficients for the log return  when market social signals are included. For most of the cryptocurrencies, the cross-correlation values are highest when considering no lag, meaning that the 24-hour social signal instantly affects the price. Adding the information about the crypto-market increases the correlation coefficient. Additionally, the correlation coefficient and the associated lags differ from one coin to another, spanning from hour delays to daily ones. This suggests both short-term and long-term impacts of social signals from news outlets and influencers, warranting further investigation in future studies.

\section{Conclusion}
In this paper, we investigated an ecosystem of 2,687 cryptocurrency influencers and 74 financial news outlets on Twitter. We have collected all tweets posted by influencers and news outlets between 10/06/2023 and 02/28/2024 and used ChatGPT and CryptoBERT to exclude irrelevant content (94\% accuracy) and detect `buy' and `not-buy' signals (84\% accuracy) from their tweets text respectively. We have also gathered time series data of the top 9 major cryptocurrency prices and explored whether (1) aggregated LLM-detected `buy' and `not-buy' signals correlate with price and (2) co-mentioned cryptocurrencies exhibit correlation in their prices.

Our exploratory results showed that (1) influencers post about 9 times as many tweets as news outlets about cryptocurrencies; 
(2) of the 470,658 tweets posted by influencers and news outlets between 10/06/2023 and 02/28/2024, 54\% were classified as implying a `buy' signal and 66\% implying a `not-buy' signal; (3) 44.4\% of all tweets posted by influencers and news outlets only contain a single coin name; and (4) tweets containing a `buy' signal typically mention at least one specific coin, whereas, ‘not-buy’ tweets often do not specify any coin name and usually focus on general market conditions.

Our time series analysis results showed that while there is a statistically significant correlation between the attributes of the co-mentioned cryptocurrencies, the effect size is small (0.17). This suggests that influencers and news outlets run other mechanisms to influence the mentioning patterns within their Twitter posts. 
With respect to the correlation between the co-mentioned cryptocurrencies and their market prices, our analyses revealed that while the correlation values for short-term dynamics consistently exceed 0.3 except for SHIB across all pairs, when examining long-term dynamics, several pairs show correlations close to zero or even negative values. Finally, our analysis of cryptocurrency prices and social signals revealed mixed results, with some cryptocurrencies showing a positive correlation with trading signals and others showing zero or negative correlations. 

Furthermore, this research limited its scope by testing only a few formulas for obtaining social signals. However, there are additional possibilities to explore. For instance, one could assign more weight to users with larger followings or those who are more central within the follower network according to centrality measures. Similarly, greater weight could be given to tweets with numerous likes and retweets. Another avenue for improvement is developing methods to identify fake accounts, bots, misinformation spreaders, and campaigns attempting to manipulate the market.Future research should study other social media platforms where cryptocurrency supporters and influencers are actively posting, such as Reddit, Telegram, and YouTube, as well as exploring mechanisms that could explain the sort of relationships we found in this paper.

\begin{credits}
\subsubsection{\ackname}We thank Atena Jafari, Daria Stetsenko, Mahdis Abbasi, Dina Della Casa, Zahra Baghshahi, Fabio Melliger, Sarvenaz Ebrahimi, and Maria Korobeynokiva for their outstanding research assistance. Yasaman Asgari thanks the University of Zurich and the Digital Society Initiative for (partially) financing this project.

\subsubsection{\discintname}
The authors declare that they have no competing interests.
\end{credits}

\printbibliography

@article{alizadeh2023open,
  title={Open-source large language models outperform crowd workers and approach ChatGPT in text-annotation tasks},
  author={Alizadeh, Meysam and Kubli, Ma{\"e}l and Samei, Zeynab and Dehghani, Shirin and Bermeo, Juan Diego and Korobeynikova, Maria and Gilardi, Fabrizio},
  journal={arXiv preprint arXiv:2307.02179},
  year={2023}
}

@article{guidi2020blockchain,
  title={When blockchain meets online social networks},
  author={Guidi, Barbara},
  journal={Pervasive and Mobile Computing},
  volume={62},
  pages={101131},
  year={2020},
  publisher={Elsevier}
}

@article{sockin2023decentralization,
  title={Decentralization through tokenization},
  author={Sockin, Michael and Xiong, Wei},
  journal={The Journal of Finance},
  volume={78},
  number={1},
  pages={247--299},
  year={2023},
  publisher={Wiley Online Library}
}

@article{kraaijeveld2020predictive,
  title={The predictive power of public Twitter sentiment for forecasting cryptocurrency prices},
  author={Kraaijeveld, Olivier and De Smedt, Johannes},
  journal={Journal of International Financial Markets, Institutions and Money},
  volume={65},
  pages={101188},
  year={2020},
  publisher={Elsevier}
}

@article{ante2023elon,
  title={How Elon Musk's twitter activity moves cryptocurrency markets},
  author={Ante, Lennart},
  journal={Technological Forecasting and Social Change},
  volume={186},
  pages={122112},
  year={2023},
  publisher={Elsevier}
}

@article{kaminski2014nowcasting,
  title={Nowcasting the bitcoin market with twitter signals},
  author={Kaminski, Jermain},
  journal={arXiv preprint arXiv:1406.7577},
  year={2014}
}

@article{kahneman2003maps,
  title={Maps of bounded rationality: Psychology for behavioral economics},
  author={Kahneman, Daniel},
  journal={American economic review},
  volume={93},
  number={5},
  pages={1449--1475},
  year={2003},
  publisher={American Economic Association}
}

@book{peterson2016trading,
  title={Trading on sentiment: The power of minds over markets},
  author={Peterson, Richard L},
  year={2016},
  publisher={John Wiley \& Sons}
}

@article{li2018more,
  title={More than just noise? Examining the information content of stock microblogs on financial markets},
  author={Li, Ting and van Dalen, Jan and van Rees, Pieter Jan},
  journal={Journal of Information Technology},
  volume={33},
  number={1},
  pages={50--69},
  year={2018},
  publisher={SAGE Publications Sage UK: London, England}
}

@article{zou2023prebit,
  title={PreBit—A multimodal model with Twitter FinBERT embeddings for extreme price movement prediction of Bitcoin},
  author={Zou, Yanzhao and Herremans, Dorien},
  journal={Expert Systems with Applications},
  volume={233},
  pages={120838},
  year={2023},
  publisher={Elsevier}
}

@inproceedings{kwak2010twitter,
  title={What is Twitter, a social network or a news media?},
  author={Kwak, Haewoon and Lee, Changhyun and Park, Hosung and Moon, Sue},
  booktitle={Proceedings of the 19th international conference on World wide web},
  pages={591--600},
  year={2010}
}

@article{gilardi2023chatgpt,
  title={ChatGPT outperforms crowd workers for text-annotation tasks},
  author={Gilardi, Fabrizio and Alizadeh, Meysam and Kubli, Ma{\"e}l},
  journal={Proceedings of the National Academy of Sciences},
  volume={120},
  number={30},
  pages={e2305016120},
  year={2023},
  publisher={National Acad Sciences}
}

@article{garcia2015social,
  title={Social signals and algorithmic trading of Bitcoin},
  author={Garcia, David and Schweitzer, Frank},
  journal={Royal Society open science},
  volume={2},
  number={9},
  pages={150288},
  year={2015},
  publisher={The Royal Society Publishing}
}

@article{mao2011predicting,
  title={Predicting financial markets: Comparing survey, news, twitter and search engine data},
  author={Mao, Huina and Counts, Scott and Bollen, Johan},
  journal={arXiv preprint arXiv:1112.1051},
  year={2011}
}

@article{karalevicius2018using,
  title={Using sentiment analysis to predict interday Bitcoin price movements},
  author={Karalevicius, Vytautas and Degrande, Niels and De Weerdt, Jochen},
  journal={The Journal of Risk Finance},
  volume={19},
  number={1},
  pages={56--75},
  year={2018},
  publisher={Emerald Publishing Limited}
}

@article{mai2015impacts,
  title={The impacts of social media on Bitcoin performance},
  author={Mai, Feng and Bai, Qing and Shan, Jay and Wang, Xin Shane and Chiang, Roger HL},
  year={2015}
}

@article{georgoula2015using,
  title={Using time-series and sentiment analysis to detect the determinants of bitcoin prices},
  author={Georgoula, Ifigeneia and Pournarakis, Demitrios and Bilanakos, Christos and Sotiropoulos, Dionisios and Giaglis, George M},
  journal={Available at SSRN 2607167},
  year={2015}
}

@article{zhang2011predicting,
  title={Predicting stock market indicators through twitter “I hope it is not as bad as I fear”},
  author={Zhang, Xue and Fuehres, Hauke and Gloor, Peter A},
  journal={Procedia-Social and Behavioral Sciences},
  volume={26},
  pages={55--62},
  year={2011},
  publisher={Elsevier}
}

@article{kim2017bitcoin,
  title={When Bitcoin encounters information in an online forum: Using text mining to analyse user opinions and predict value fluctuation},
  author={Kim, Young Bin and Lee, Jurim and Park, Nuri and Choo, Jaegul and Kim, Jong-Hyun and Kim, Chang Hun},
  journal={PloS one},
  volume={12},
  number={5},
  pages={e0177630},
  year={2017},
  publisher={Public Library of Science San Francisco, CA USA}
}

@article{fama1969adjustment,
  title={The adjustment of stock prices to new information},
  author={Fama, Eugene F and Fisher, Lawrence and Jensen, Michael C and Roll, Richard},
  journal={International economic review},
  volume={10},
  number={1},
  pages={1--21},
  year={1969},
  publisher={JSTOR}
}

@article{li2014news,
  title={News impact on stock price return via sentiment analysis},
  author={Li, Xiaodong and Xie, Haoran and Chen, Li and Wang, Jianping and Deng, Xiaotie},
  journal={Knowledge-Based Systems},
  volume={69},
  pages={14--23},
  year={2014},
  publisher={Elsevier}
}

@article{bollen2011twitter,
  title={Twitter mood predicts the stock market},
  author={Bollen, Johan and Mao, Huina and Zeng, Xiaojun},
  journal={Journal of computational science},
  volume={2},
  number={1},
  pages={1--8},
  year={2011},
  publisher={Elsevier}
}

@article{kulakowski2023sentiment,
  title={Sentiment Classification of Cryptocurrency-Related Social Media Posts},
  author={Kulakowski, Mikolaj and Frasincar, Flavius},
  journal={IEEE Intelligent Systems},
  volume={38},
  number={4},
  pages={5--9},
  year={2023},
  publisher={IEEE}
}

@article{kaplanski2010sentiment,
  title={Sentiment and stock prices: The case of aviation disasters},
  author={Kaplanski, Guy and Levy, Haim},
  journal={Journal of financial economics},
  volume={95},
  number={2},
  pages={174--201},
  year={2010},
  publisher={Elsevier}
}

@article{baker2007investor,
  title={Investor sentiment in the stock market},
  author={Baker, Malcolm and Wurgler, Jeffrey},
  journal={Journal of economic perspectives},
  volume={21},
  number={2},
  pages={129--151},
  year={2007},
  publisher={American Economic Association}
}

@article{valle2022does,
  title={Does twitter affect stock market decisions? financial sentiment analysis during pandemics: A comparative study of the h1n1 and the covid-19 periods},
  author={Valle-Cruz, David and Fernandez-Cortez, Vanessa and L{\'o}pez-Chau, Asdr{\'u}bal and Sandoval-Almaz{\'a}n, Rodrigo},
  journal={Cognitive computation},
  volume={14},
  number={1},
  pages={372--387},
  year={2022},
  publisher={Springer}
}

@article{shen2019does,
  title={Does twitter predict Bitcoin?},
  author={Shen, Dehua and Urquhart, Andrew and Wang, Pengfei},
  journal={Economics letters},
  volume={174},
  pages={118--122},
  year={2019},
  publisher={Elsevier}
}

@article{cano2023twitter,
  title={Twitter as a predictive system: a systematic literature review},
  author={Cano-Marin, Enrique and Mora-Cantallops, Mar{\c{c}}al and S{\'a}nchez-Alonso, Salvador},
  journal={Journal of Business Research},
  volume={157},
  pages={113561},
  year={2023},
  publisher={Elsevier}
}

@article{mungo2024cryptocurrency,
  title={Cryptocurrency co-investment network: token returns reflect investment patterns},
  author={Mungo, Luca and Bartolucci, Silvia and Alessandretti, Laura},
  journal={EPJ Data Science},
  volume={13},
  number={1},
  pages={11},
  year={2024},
  publisher={Springer Berlin Heidelberg}
}
\end{document}